\documentclass[12pt,twoside]{article}
 \usepackage{amsmath}
 \usepackage{amssymb}
 \usepackage{array}
 \usepackage{graphicx}
 \usepackage{geometry}
\usepackage{graphics}
 \usepackage{enumerate}
 \usepackage[utf8]{inputenc}
\begin{document} \vskip3cm\noindent
 \setcounter{page}{1}
 \font\tinyfont=cmr8 \font\headd=cmr8
 \pagestyle{myheadings}
 \begin{center}
 \vskip.2cm\noindent {\bf On Extended d-D Kappa Distribution}\end{center}
 \vskip.5cm\noindent\begin{center} Arak M. Mathai\\
 \vskip.2cm\noindent Department of Mathematics and Statistics, McGill University,\\ 
Montreal, Canada\\
 a.mathai@mcgill.ca\\
 \vskip.2cm\noindent and\\
 \vskip.2cm\noindent Hans J. Haubold\\
 Office for Outer  Space Affairs, United Nations, Vienna International Centre, Vienna, Austria\\
 hans.haubold@gmail.com\end{center}
 \vskip.3cm\noindent{\bf Abstract.}
 Thermal Doppler broadening of spectral profiles for particle populations in the absence or presence of potential fields are described by kappa distributions. The kappa distribution provides a replacement for the Maxwell-Boltzmann distribution which can be considered as a generalization for describing systems characterized by local correlations among their particles as found in space and astrophysical plasmas. This paper presents all special cases of kappa distributions as members of a general pathway family of densities introduced by Mathai.
\vskip.3cm\noindent{\bf Key words:}
atomic data - line: profiles - methods: analytical - plasmas 
 \vskip.5cm

  \vskip.5cm\noindent{\bf 1.\hskip.3cm Introduction}
  \vskip.3cm
  A particle is in thermal equilibrium when the exchange of heat or entropy in the system stops. The kappa index controls the exchange of entropy and it is also a measure of departure from the equilibrium state. The particle velocity distribution can be given in terms of a kappa distribution. In the case of a collisional plasma, where no local correlations among particles exist, the system is stabilized into a Maxwell-Boltzmann distribution. The kappa index is inversely proportional to the correlation between the energies of two particles.
   The d-D kappa distribution describes the particle velocity with dimensionality $d$. Livadiotis (2018), Livadiotis and McComas (2023), and Cuesta (2024) take this particle velocity density as the following:
   $$f(X;\theta,k_d)=c_d[1+\frac{1}{k_d-\frac{d}{2}}\frac{(X-\mu)'(X-\mu)}{\theta^2}]^{-(k_d+1)}\eqno(1.1)
   $$where $X$ is a $d\times 1$ velocity vector, a prime denotes the transpose and $\mu=E[X]=<X>$ is the expected value of $X$, which is also a location parameter vector, where $E[(\cdot)]$ denotes the expected value with respect to the density $f(X;\theta,k_d)$, and the normalizing constant $c_d$ is the following:
   $$c_d=\frac{\Gamma(k_d+1)}{\Gamma(k_d-\frac{d}{2}+1)[\pi(k_d-\frac{d}{2})\theta^2]^{\frac{d}{2}}}\eqno(1.2)
   $$and $\theta=\sqrt{k_BT/m}$ is the thermal speed of the particle with mass $m$ and temperature $T$ expressed in speed units. If $k_d-\frac{d}{2}$ is invariant over the dimensionality $d$ then let $k_o$ be this constant value. Then, $k_d=k_o+\frac{d}{2}$. Then, the normalizing constant becomes
   $$c_d=\frac{\Gamma(k_o+1+\frac{d}{2})}{\Gamma(k_o+1)[\pi k_o\theta^2]^{\frac{d}{2}}}\eqno(1.3)
   $$and the functional part is
   $$[1+\frac{1}{k_d-\frac{d}{2}}\frac{(X-\mu)'(X-\mu)}{\theta^2}]^{-(k_d+1)}=[1+\frac{1}{k_o}\frac{(X-\mu)'(X-\mu)}{\theta^2}]^{-(k_o+1+\frac{d}{2})}\eqno(1.4)
   $$Since $k_d-\frac{d}{2}=k_o$ is invariant $k_3-\frac{3}{2}=k_d-\frac{d}{2}$,      $k_3=k_d-\frac{(d-3)}{2}$. If $\theta$ depends on the kappa index, then $\theta$ is of the form $\theta_k$. Two forms of $\theta_k$ are usually taken. One is $\theta_k^2=\frac{(k_3-\frac{3}{2})}{k_3}\frac{(d-1)}{2}\theta^2$ and the other is $\theta_{k,d}^2=\frac{k_o}{k_d}\theta^2=\frac{(k_d-\frac{d}{2})}{k_d}\theta^2$. If $\theta_k$ is used then (1.4) becomes
   $$[1+\frac{1}{k_d}\frac{(X-\mu)'(X-\mu)}{\theta_k^2}]^{-(k_d+\frac{1}{2}(d-1))}.\eqno(1.5)
   $$On the other hand, if $\theta_{k,d}$ is used then (1.4) becomes
   $$[1+\frac{1}{k_d}\frac{(X-\mu)'(X-\mu)}{\theta_{k,d}^2}]^{-(k_d+1)}.\eqno(1.6)
   $$Thus, one can write the kappa density in (1.1) in a number of different ways. One set of representations will be for $d=1,2,3$, another set will be with $\theta$ replaced by $\theta_k$ or $\theta_{k,d}$. When $k_o\to\infty$, (1.1) will go to the density
   $$f_1(X;\theta,k_d)=c_{*}{\rm e}^{-\frac{(X-\mu)'(X-\mu)}{\theta^2}}\eqno(1.7)
   $$where $c_{*}=(\pi \theta^2)^{-{\frac{d}{2}}}$. This density (1.6) is a multivariate real Gaussian and also becomes as $d-D$ Maxwellian distribution.

   \vskip.3cm\noindent{\bf 2.\hskip.3cm A Pathway Family of Densities}
   \vskip.3cm
   All the forms of the kappa distribution considered in Section 1 are special cases of a general pathway family of densities introduced in Mathai (2005). Let $X$ be a $p\times 1$ real vector random variable with the covariance matrix $\Sigma={\rm Cov}(X)=E[(X-E(X))(X-E(X))']$ where a prime denotes the transpose and $E(\cdot)$ is the expected value of $(\cdot)=<(\cdot)>$ with respect to the density of $X$. The $p$ components of $X$ may be correlated among themselves. One can get rid of the effect of correlations by taking $Y=\Sigma^{-\frac{1}{2}}(X-\mu)$ where $\Sigma^{\frac{1}{2}}$ is the real positive definite square root of the positive definite matrix $\Sigma>O$. Then the covariance matrix in $Y$ is the identity matrix, free of all correlations. The Euclidean distance of $Y$ from the origin is also $(X-\mu)'\Sigma^{-1}(X-\mu)$, $\mu=E(X)$. Also, $(X-\mu)'\Sigma^{-1}(X-\mu)= \mbox{ a positive constant }$ is an offset ellipsoid and it is also known in statistical literature as the ellipsoid of concentration. If the components of $X$ are to be simply weighted, rather than removing the effect of correlations, then one can replace $\Sigma^{-1}$ in the quadratic form by a real positive definite matrix $A>O$ and then we may consider $(X-\mu)'A(X-\mu)$. Consider the following density:
   $$g(X){\rm d}X=c[(X-\mu)'\Sigma^{-1}(X-\mu)]^{\gamma}[1+a((X-\mu)'\Sigma^{-1}(X-\mu))^{\delta}]^{-\eta}{\rm d}X\eqno(2.1)
   $$for $a>0,\delta>0,\eta>0,\gamma>0$, all real scalar parameters, and $c$ is the normalizing constant, which can be evaluated as
   $$c=\frac{\delta \Gamma(\frac{p}{\delta})\Gamma(\eta)a^{\frac{\gamma}{\delta}+\frac{p}{2\delta}}}{|\Sigma|^{\frac{1}{2}}\pi^{\frac{p}{2}}\Gamma(\frac{\gamma}{\delta}+\frac{p}{2\delta})\Gamma(\eta-\frac{\gamma}{\delta}-\frac{p}{2\delta})},
    \eta-\frac{\gamma}{\delta}-\frac{p}{2\delta}>0\eqno(2.2)
   $$see the derivation of $c$ in the Appendix. Consider $\delta=1,\Sigma=I$ (identity matrix), $a=\frac{1}{\theta^2(k_d-\frac{d}{2})},\eta=k_d+1,\gamma=0$. Then (2.1) reduces to (1.1), the density given by Livadiotis (2018). The general representation in (2.1) and (2.2) has many advantages. This (2.1) is a member of the pathway family of distributions defined in Mathai (2005). Let $a=\frac{b}{\alpha-a_o},b>0$, $\alpha>a_o$ and $\eta=\rho(\alpha-a_o),\rho>0$. Then, (2.1) becomes
   $$g_1(X){\rm d}X=c_1[(X-\mu)'\Sigma^{-1}(X-\mu)]^{\gamma}[1+\frac{b}{\alpha-a_o}((X-\mu)'\Sigma^{-1}(X-\mu))^{\delta}]^{-\rho(\alpha-a_0)}{\rm d}X.\eqno(2.3)
   $$Note that when $\alpha\to a_o$ from the right, $g_1(X)$ goes to the density
   $$g_2(X){\rm d}X=c_2[(X-\mu)'\Sigma^{-1}(X-\mu)]^{\gamma}{\rm e}^{-b\rho[(X-\mu)'\Sigma^{-1}(X-\mu)]^{\delta}}{\rm d}X\eqno(2.4)
   $$where
   $$c_2=\frac{\delta\Gamma(\frac{p}{2})(b\rho)^{\frac{\gamma}{\delta}+\frac{p}{2\delta}}}{\pi^{\frac{p}{2}}|\Sigma|^{\frac{1}{2}}\Gamma(\frac{\gamma}{\delta}+\frac{p}{2\delta})},\delta>0,b>0,\rho>0.\eqno(2.5)
   $$The density in (2.4) can be taken as a power-transformed real multivariate Maxwell-Boltzmann density. For $\delta=1$, (2.4) is a form of multivariate Maxwell-Botlzmann density. Hence, if (2.4) is the stable density in a physical system, then the unstable neighborhood and transitional stages are determined by (2.3) for various values of the pathway parameter $\alpha$. We may also note that for $\alpha=k_d,a_o=\frac{d}{2},b=\frac{1}{\theta^2},\delta=1,\gamma=0,\rho=\frac{k_d+1}{k_d-\frac{d}{2}}$ one has (1.1) also. This density in (2.3) has another advantage. For $\alpha<a_o$ we can write $\alpha-a_o=-(a_o-\alpha),a_o-\alpha>0$ so that the density in (2.3) switches into a type-1 beta form of the density given by
   $$g_3(X){\rm d }X=c_3[(X-\mu)'\Sigma^{-1}(X-\mu)]^{\gamma}[1-\frac{b}{a_o-\alpha}((X-\mu)'\Sigma^{-1}(X_\mu))^{\delta}]^{\rho(a_o-\alpha)}{\rm d}X,\alpha<a_o\eqno(2.6)
   $$where $[1-\frac{b}{a_o-\alpha}((X-\mu)'\Sigma^{-1}(X-\mu))^{\delta}]>0$ so that the density is defined within the ellipsoid
   $$(X-\mu)'\Sigma^{-1}(X-\mu)=(\frac{a_o-\alpha}{b})^{\frac{1}{\delta}}\mbox{  for  }\alpha<a_o,b>0,\delta>0
   $$and
   $$c_3=\frac{\delta\Gamma(\frac{p}{\delta})\Gamma(1+\rho(a_o-\alpha)+\frac{\gamma}{\delta}+\frac{p}{2\delta})(\frac{b}{a_o-\alpha})^{\frac{\gamma}{\delta}+\frac{p}{2\delta}}}{|\Sigma|^{\frac{1}{2}}\pi^{\frac{p}{2}}\Gamma(\frac{\gamma}{\delta}+\frac{p}{2\delta})\Gamma(1+\rho(a_o-\alpha))},\alpha<a_o.\eqno(2.7)
   $$This normalizing constant is evaluated by going through the procedure in Appendix A1 and the final integral is evaluated by using a type-1 beta integral. Thus, $g_1,g_2,g_3$ belong to the pathway family of densities. If the power-transformed Maxwell-Boltzmann density in (2.4) is the stable density in a physical system then the unstable neighborhoods and the transitional stages are given by $g_1$ in (2.3) and $g_3$ in (2.6). One can switch among a generalized type-1 beta family, a type-2 beta family and a gamma family or Maxwell-Boltzmann family of distributions through the pathway parameter $\alpha$. In the model in (2.3) one can identify $\alpha$ with $k_d$ and $a_o$ with $\frac{d}{2}$ if convenient.
   \vskip.3cm\noindent{\bf 3.\hskip.3cm Livadiotis' $d-D$ Density Through an entropy Optimization}
   \vskip.3cm
   Let $X$ be a $p\times 1$ real vector random variable. Let $f(X)$ be a density function, that is, $f(X)\ge 0$ for all $X$ and $\int_Xf(X){\rm d}X=1$ where $f(X)$ is a real-valued scalar function of $X$. Consider Mathai's entropy for the vector random variable, namely,
   $$M_{\alpha}(f)=\frac{\int_X[f(X)]^{1+\frac{a_o-\alpha}{\eta}}{\rm d}X-1}{\alpha-a_o}\eqno(3.1)
   $$for $a_o$ a fixed quantity or anchoring point, $\alpha$ is the pathway parameter and the deviation of $\alpha$ from $a_o$ is measured in $\eta>0$ units. Then, when $\alpha\to a_o$, we can see that (3.1) reduces to Shannon's entropy $S(f)=-K\int_Xf(X)\ln f(X){\rm d}X$ where $K$ is a constant. Shannon's entropy is for the scalar variable case and the corresponding real vector-variate form is denoted here as $S(f)$. Let $\Sigma>O$ be the covariance matrix of $X$. Consider the ellipsoid of concentration $(X-\mu)'\Sigma^{-1}(X_\mu)=$ a positive constant, where $\mu$ is a $p\times 1$ location parameter vector. We will set moment-type constraints on the ellipsoid of concentration and then optimize (3.1). Consider the following constraints:
   $$E[(X-\mu)'\Sigma^{-1}(X-\mu)]^{\gamma(\frac{a_o-\alpha}{\eta})}=\mbox{ fixed }, E[(X-\mu)'\Sigma^{-1}(X-\mu)]^{\gamma(\frac{a_o-\alpha}{\eta})+\delta}=\mbox{ fixed}\eqno(3.2)
   $$for some parameters $\gamma>0,\eta>0,\delta>0$. If we use calculus of variation for the optimization, then the Euler equation becomes the following where $\lambda_1$ and $\lambda_2$ are Lagrangian multipliers:
   $$\frac{\partial}{\partial f}[f^{1+\frac{a_o-\alpha}{\eta}}-\lambda_1[(X-\mu)'\Sigma^{-1}(X-\mu)]^{\gamma(\frac{a_o-\alpha}{\eta})}f+\lambda_2[(X-\mu)'\Sigma^{-1}(X-\mu)]^{\gamma(\frac{a_o-\alpha}{\eta})+\delta}f]=0.
   $$This gives the solution for $f$ as the following:
   $$f=\nu_1[(X-\mu)'\Sigma^{-1}(X-\mu)]^{\gamma}[1-\frac{\lambda_2}{\lambda_1}[(X-\mu)'\Sigma^{-1}(X-\mu)]^{\delta}]^{\frac{\eta}{a_o-\alpha}}\eqno(3.3)
   $$where $\nu_1,\lambda_1,\lambda_2$ are constants. Let $\frac{\lambda_2}{\lambda_1}=b(a_o-\alpha), b>0$ and let $\nu_1$ be the normalizing constant to make (3.3) a density. Then, for $u=(X-\mu)'\Sigma^{-1}(X-\mu)$ we have the following densities from (3.3):
   \begin{align*}
   f_1(X)&=\nu_1 u^{\gamma}[1-b(a_o-\alpha)u^{\delta}]^{\frac{\eta}{a_0-\alpha}},\alpha<a_o\tag{3.4}\\
   f_2(X)&=\nu_2 u^{\gamma}[1+b(\alpha-a_o)u^{\delta}]^{-\frac{\eta}{\alpha-a_o}},\alpha>a_o\tag{3.5}\\
   f_3(X)&=\nu_3 u^{\gamma}{\rm e}^{-b\eta u^{\delta}},\alpha\to a_o\tag{3.6}\end{align*}
   where in (3.4) we need an additional condition $1-b(a_o-\alpha)u^{\delta}>0$ in order to make (3.4) a density. Note that in the limiting form we have the following properties:
   \begin{align*}
   {\rm e}^{-b\eta u^{\delta}}&=\lim_{\alpha\to a_o}[1-b(a_o-\alpha)u^{\delta}]^{\frac{\eta}{a_o-\alpha}}=\lim_{\alpha\to a_o}[1+b(\alpha-a_o)u^{\delta}]^{-\frac{\eta}{\alpha-a_o}}\\
   &=\lim_{\alpha\to a_o}[1-\frac{b}{a_o-\alpha}u^{\delta}]^{\eta (a_o-\alpha)}=\lim_{\alpha\to a_o}[1+\frac{b}{\alpha-a_o}u^{\delta}]^{-\eta(\alpha-a_o)}\tag{3.7}\end{align*}Hence, one can take any one of the formats in (3.7) in the limiting case. Livadiotis' density in (1.1) is available from (3.5) by taking $b(\alpha-a_o)=\frac{1}{\theta(k_d-\frac{d}{2})}$ and $\frac{\eta}{\alpha-a_o}=k_d+1$.
   \vskip.2cm For $\gamma=0,\delta=1,\eta=1,a_o=1,b=1$, equations (3.4),(3.5)(3.6) give a real multivariate version of Tsallis statistics of non-extensive statistical mechanics. For $\delta=1,\eta=1,a_o=1,b=1,\delta=1$, (3.5) and (3.6) can be taken as a multivariate extension of superstatistics. Matrix-variate versions of Tsallis statistics and superstatistics can also be defined.

   \vskip.3cm\noindent{\bf 4.\hskip.3cm A Matrix-variate generalization of Livadiotis' Density}
   \vskip.3cm
   Let $Y=(y_{ij})$ be a $p\times q,p\le q$ and of rank $p$ matrix with distinct real scalar variables $y_{ij}$'s as elements. Let $g(Y)$ be a real-valued scalar function of $Y$ such that $g(Y)\ge 0$ for all $Y$ and $\int_Yg(Y){\rm d}Y=1$ so that $g(Y)$ is a density function where ${\rm d}Y=\wedge_{i=1}^p\wedge_{j=1}^q{\rm d}y_{ij}=$ the wedge product of all distinct differentials in $Y$. Let $M$ be a $p\times q,p\le q$ parameter matrix. Let $A>O$ be $p\times p$ and $B>O$ be $q\times q$ positive definite constant matrices. Let $A^{\frac{1}{2}}$ and $B^{\frac{1}{2}}$ be the positive definite square roots of $A$ and $B$ respectively. Let
   $$U=A^{\frac{1}{2}}(Y-M)B^{\frac{1}{2}}, V=UU'=A^{\frac{1}{2}}(Y-M)B(Y-M)'A^{\frac{1}{2}}\eqno(4.1)
   $$Then, the determinant of $V$, namely $|V|$, can be interpreted in different ways. $|V|$ is the product of the eigenvalues of the matrix $V$. Let $U_1,...,U_p$ be the $p$ linearly independent rows of $U$, the $p\times q,p\le q$ and of rank $p$ matrix $U$. Then, $U_j,j=1,...,p$ can be taken as $p$ linearly independent points in a $q$-dimensional Euclidean space with $p\le q$. Then, $|V|^{\frac{1}{2}}=|UU'|^{\frac{1}{2}}=$ the volume of the $p$-parallelotope generated in the convex hull of the $p$ linearly independent points, taken in the order. Thus, $|V|$ is also the square of the volume content of this parallelotope. Consider Mathis's entropy in (3.1) with $f(X)$ replaced by $g(Y)$ where $Y$ is now $p\times q,p\le q$ matrix of rank $p$. Consider the optimization of (3.1) with the real-valued scalar function $g(Y)$ under the following constraints:
   $$E[|V|^{\gamma(\frac{a_o-\alpha}{\eta})}=\mbox{ fixed }, E[|V|^{\gamma(\frac{a_o-\alpha}{\eta})}[{\rm tr}(V)]^{\delta}=\mbox{ fixed}\eqno(4.2)
   $$Then, proceeding as in Section 3, we can end up with the following pathway densities:
   \begin{align*}
   g_1(Y)&=C_1|V|^{\gamma}[1-b(a_o-\alpha)({\rm tr}(V))]^{\frac{\eta}{a_o-\alpha}},\alpha<a_o\tag{4.3}\\
   g_+2(Y)&=C_2|V|^{\gamma}[1+b(\alpha-a_o)({\rm tr}(V))]^{-\frac{\eta}{\alpha-a_o}},\alpha>a_o\tag{4.4}\\
   g_3(Y)&=|V|^{\gamma}{\rm e}^{-b\eta {\rm tr}(V)}\tag{4.5}\end{align*}
   where $C_1,C_2,C_3$ are the normalizing constants and in (4.3) the additional condition needed is $1-b(a_o-\alpha){\rm tr}(V)>0$ to make (4.3) a density, where
   $$V=A^{\frac{1}{2}}(Y-M)B(Y-M)'A^{\frac{1}{2}}.\eqno(4.6)
   $$The normalizing constants are evaluated in Appendix A2 by using the following steps:
   $$U=A^{\frac{1}{2}}(Y-M)B^{\frac{1}{2}}\Rightarrow {\rm d}U=|A|^{\frac{q}{2}}|B|^{\frac{p}{2}}{\rm d}Y
   $$and
   $$V=UU'\Rightarrow {\rm d}U=\frac{\pi^{\frac{pq}{2}}}{\Gamma_p(\frac{q}{2})}|V|^{\frac{q}{2}-\frac{p+1}{2}}{\rm d}V\eqno(i)
   $$where $\Gamma_p(\alpha)$ is a real matrix-variate gamma function, associated with a real matrix-variate gamma integral, and it is the following:
    \begin{align*}
    \Gamma_p(\alpha)&=\int_{S>O}|S|^{\alpha-\frac{p+1}{2}}{\rm e}^{-{\rm tr}(S)}{\rm d}S,\Re(\alpha)>\frac{p-1}{2}\tag{ii}\\
    &=\pi^{\frac{p(p-1)}{4}}\Gamma(\alpha)\Gamma(\alpha-\frac{1}{2})...\Gamma(\alpha-\frac{p-1}{2}),\Re(\alpha)>\frac{p-1}{2}\tag{iii}\end{align*}where $\Re(\cdot)$ denotes the real part of $(\cdot)$. Note that (4.5) is a real rectangular matrix-variate Maxwell-Boltzmann density.

   \vskip.3cm\noindent{\bf Note 4.1.}\hskip.3cm Results parallel to all the results in Sections 3 and 4 are also available in the complex domain. Corresponding physics can also be dealt with in the complex domain. A real rectangular matrix-variate version of Tsallis statistics is available from (4.3)-(4.5) for $\gamma+\frac{q}{2}=\frac{p+1}{2},b=1,\alpha=1,a_o=1,\eta=1$. Corresponding results in the complex domain can also be worked out. For $b=1,\alpha=1,\eta=1$, (4.4) and (4.5) give a real rectangular matrix-variate version of superstatistics. Corresponding versions in the complex domain can also be worked out. Such rectangular or square matrix-variate versions may not be available in the literature.

   \vskip.3cm\begin{center}{\bf APPENDIX}\end{center}
   \vskip.3cm\noindent{A1: Derivation of the normalizing constant in (2.2)}

   \vskip.3cm Let $X$ be a $p\times 1$ vector random variable and $\mu=E[X], \Sigma={\rm Cov}(X)$.
   $$\frac{1}{c}=\int_X[(X-\mu)'\Sigma^{-1}(X-\mu)]^{\gamma}[1+a((X-\mu)'\Sigma^{-1}(X-\mu))^{\delta}]^{-\eta}{\rm d}X.
   $$Consider the transformation $$X=\Sigma^{-\frac{1}{2}}(X-\mu)\Rightarrow {\rm d}Z=|\Sigma|^{-\frac{1}{2}}{\rm d}X
   $$see Mathai (1997). Let $u=ZZ'$. Then writing $Z$ uniquely as a product of a unique lower triangular matrix and a unique semi-orthonormal matrix and then integrating out the differential element over a Stiefel manifold, we have a relation between ${\rm d}u$ and ${\rm d}Z$, that is
   $${\rm d}Z=\frac{\pi^{\frac{q}{2}}}{\Gamma(\frac{p}{2})}u^{\frac{p}{2}-1}{\rm d}u,u>0\eqno(A_1)
   $$see Mathai (1997) for the details. Observe that $u$ is real scalar whereas $Z$ is a $p\times 1$ vector. Then
   \begin{align*}
   &\int_X[(X-\mu)'\Sigma^{-1}(X-\mu)]^{\gamma}[1+a((X-\mu)'\Sigma^{-1}(X-\mu))^{\delta}]^{-\eta}{\rm d}X\\
   &=\frac{|\Sigma|^{\frac{1}{2}}\pi^{\frac{p}{2}}}{\Gamma(\frac{p}{2})}\int_{u=0}^{\infty}u^{\gamma+\frac{p}{2}-1}(1+au^{\delta})^{-\eta}{\rm d}u.\end{align*}
   Now, let $v=u^{\delta},u>0,\delta>0\Rightarrow {\rm d}u=\frac{1}{\delta}v^{\frac{1}{\delta}-1}{\rm d}v$. Integrating out by using a type-2 beta integral we have
   $$\frac{1}{c}=\frac{|\Sigma|^{\frac{1}{2}}\pi^{\frac{p}{2}}}{ \Gamma(\frac{p}{2})}\frac{\Gamma(\frac{\gamma}{\delta}+\frac{p}{2\delta})\Gamma(\eta-\frac{\gamma}{\delta}-\frac{p}{2\delta})}{\delta a^{\frac{\gamma}{\delta}+\frac{p}{2\delta}}\Gamma(\eta)},\eta-\frac{\gamma}{\delta}-\frac{p}{2\delta}>0\eqno(A_2)
   $$which gives the normalizing constant.

   \vskip.3cm\noindent {\bf A2\hskip.3cm Derivation of the normalizing constant $C$ in (4.4)}
   \vskip.3cm\noindent Let $Y$ be $p\times q,p\le q$ and of rank $p$ matrix of real scalar random variables as elements.
   $$\frac{1}{C}=\int_Y[A^{\frac{1}{2}}(Y-M)B(Y-M)'A^{\frac{1}{2}}]^{\gamma}[1+b(\alpha-a_o){\rm tr}[(A^{\frac{1}{2}}(Y-M)B(Y-M)'A^{\frac{1}{2}})]^{-\frac{\eta}{\alpha-a_o}}{\rm d}Y,
   $$for $\alpha>a_o.$ Let $U=A^{\frac{1}{2}}(Y-M)B^{\frac{1}{2}}\Rightarrow {\rm d}U=|A|^{\frac{q}{2}}|B|^{\frac{p}{2}}{\rm d}Y$ where $Y$ is $p\times q,p\le q$ and of rank $p$, $M$ is a $p\times q,p\le q$ parameter matrix, $A>O,B>O$ are $p\times p$ and $q\times q$ constant positive definite matrices, see Mathai (1997) for the Jacobian in this transformation. Let $V=UU'\Rightarrow {\rm d}U=\frac{\pi^{\frac{pq}{2}}}{\Gamma_p(\frac{q}{2})}|V|^{\frac{q}{2}-\frac{p+1}{2}}{\rm d}V$, see Mathai (1997) for details. Then

   \begin{align*}
   \frac{1}{C}&=|A|^{-\frac{q}{2}}|B|^{-\frac{p}{2}}\frac{\pi^{\frac{pq}{2}}}{\Gamma_p(\frac{q}{2})}\int_V|V|^{\gamma+\frac{q}{2}-\frac{p+1}{2}}\\
   &\times [1+b(\alpha-a){\rm tr}(V)]^{-\frac{\eta}{\alpha-a_o}}{\rm d}V,\alpha>a_o\end{align*}

   We can replace one factor by an equivalent integral, that is,

   $$[1+b(\alpha-a_o){\rm tr}(V)]^{-\frac{\eta}{\alpha-a_o}}\equiv \frac{1}{\Gamma(\frac{\eta}{\alpha-a_o})}\int_{z=0}^{\infty}z^{\frac{\eta}{\alpha-a_o}-1}{\rm e}^{-z(1+b(\alpha-a_o){\rm tr}(V))}{\rm d}z.$$

   $$\frac{1}{C}=|A|^{-\frac{q}{2}}|B|^{-\frac{p}{2}}\frac{\pi^{\frac{pq}{2}}}{\Gamma_p(\frac{q}{2})}{\rm e}^{-z}[\int_V|V|^{\gamma+\frac{q}{2}-\frac{p+1}{2}}{\rm e}^{-z(\alpha-a_o){\rm tr}(V)}]{\rm d}V $$Evaluate the V-integral by using a real matrix-variate gamma integral. We have

   \begin{align*}
   \frac{1}{C}&=|A|^{-\frac{q}{2}}|B|^{-\frac{p}{2}}\frac{\pi^{\frac{pq}{2}}}{\Gamma_p(\frac{q}{2})}[b(\alpha-a_o)]^{-p(\gamma+\frac{q}{2})}\Gamma_p(\gamma+\frac{q}{2})\\
   &\times\int_{z=0}^{\infty}z^{\frac{\eta}{\alpha-a_o}-p(\gamma+\frac{q}{2})-1}{\rm e}^{-z}{\rm d}z\\ &=|A|^{-\frac{q}{2}}|B|^{-\frac{p}{2}}\frac{\pi^{\frac{pq}{2}}}{\Gamma_p(\frac{q}{2})}\Gamma_p(\gamma+\frac{q}{2})[b(\alpha-a_o)]^{-p(\gamma+\frac{q}{2})}\Gamma(\frac{\eta}{\alpha-a_o}-p(\gamma+\frac{q}{2}))\end{align*} for $\frac{\eta}{\alpha-a_o}-p(\gamma+\frac{q}{2})>0$. This completes the computations.

   \vskip.3cm\begin{center}{\bf References}\end{center}
   \vskip.3cm\noindent M.E. Cuesta, A.T. Cummings, G. Livadiotis, D.J. McComas, C.M.S. Cohen, L.Y. Khoo, T. Sharma, M.M. Shen, R. Bandyopadhyay, J.S. Rankins, J.R. Szalay, H.A. Farooki, Z. Xu, G.D. Muro, M.L. Stevens, and S.D. Bale (2024): Observations of Kappa Distributions in Solar Energetic Protons and Derived Thermodynamic Properties, https://arxiv.org/abs/2407.20343.
   \vskip.3cm\noindent G. Livadiotis (2018): Thermal Doppler broadening of spectral emissions in space and astrophysical plasmas. {\it The Astrophysical Journal Supplement Series}, {\bf 239:25}, https://doi.org/10.3847/1538-4365/aae835.
   \vskip.3cm\noindent G. Livadiotis and D.J. McComas (2023): Entropy defect in thermodynamics. {\it scientific reports}, {\bf 13:9033}, https://doi.org/10.1038/s41598-023-36080-w.
   \vskip.2cm\noindent A.M. Mathai (1997): {\it Jacobians of Matrix Transformations and Functions of Matrix Argument}, World Scientific Publishing, New York.
   \vskip.2cm\noindent A.M. Mathai (2005): A pathway to matrix-variate gamma and normal densities. {\it Linear Algebra and its Applications}, {\bf 396}, 317-328.

   \end{document}